\documentclass[prd,aps,superscriptaddress,preprintnumbers,notitlepage,nofootinbib]{revtex4}
\usepackage{tablefootnote}
\usepackage[T1]{fontenc}
\usepackage{amsmath}
\usepackage{amssymb}
\usepackage{adjustbox}
\usepackage{multirow}
\usepackage{hyperref}
\usepackage{bbm}
\usepackage{xcolor}
\usepackage{braket}
\usepackage{slashed}
\usepackage{diagbox}
\usepackage[utf8]{inputenc}
\usepackage{tcolorbox}
\usepackage{tabularx}
\usepackage{bbm,bm}

\newcommand{\p}[1]{$\mathbf{P}$}
\newcommand{\cp}[1]{$\mathbf{CP}$}

\begin{document}
\raggedbottom

\title{Modified Tri-bimaximal neutrino mixing confronted by JUNO $\theta_{12}$ measurement}

\author{Xiao-Gang He}

\affiliation{State Key Laboratory of Dark Matter Physics, Tsung-Dao Lee Institute, Shanghai Jiao Tong University, Shanghai 200240, China} 
\affiliation{Key Laboratory for Particle Astrophysics and Cosmology (MOE) \& Shanghai Key Laboratory for Particle Physics and Cosmology, Shanghai Jiao Tong University, Shanghai 200240, China}

\date{\today}

\begin{abstract}
The JUNO collaboration has released its first measurement of reactor neutrino oscillations, obtaining $\sin\theta^2_{12} = 0.3092 \pm 0.0087$, an improvement inprecision by a factor of 1.6 over previous combined results. We confront the minimally modifred tri-bimaximal mixing pattern with the new data. 
Before the measurement of a nonzero $\theta_{13}$ mixing angle, the tri-bimximal mixing pattern is one of the most popular simple mixing scheme for neutrinos. 
Modifications have been proposed to keep some features of tri-bimaximal mixing and make it to be consistent with data. 
Minimal modifications preserving one column of the trim-bimaximal mixing matrix, yielding three patterns: a) unchanged third, b) unchanged second, and c) unchanged first column. Pattern a) is excluded since it keeps $\theta_{13}=0$. Pattern b) predicts $V_{e2}=1/\sqrt{3}$ and specific CP-phase correlations, but JUNO’s smaller $|V_{e2}|$ disfavors it at more than $3.5\sigma$. Pattern c), predicting $V_{e2}=\cos\tau/\sqrt{3}$, can be in agreement with current data within $1\sigma$, and implies $\sin^2\theta_{12}=(1-3\sin^2\theta_{13})/3(1-\sin^2_{13})$ and CP violating quantity $\sin\delta_{CP}= \pm 0.998$. The negtive sign favors the inverted neutrino mass hierarchy. Upcoming experiments can further test this scenario.
\end{abstract}

\maketitle

%%%%%%%%%%%%%%%%%%%%%
%%%%%%%%%%%%%%%%%%%%%
Neutrino mixing properties are fundamentally important in particle physics~\cite{PMNS, PDG}. The existence of neutrino mixing implies that at least two neutrinos are massive, providing clear evidence for physics beyond the Standard Model (SM). Tremendous efforts have been devoted to understanding the neutrino mixing pattern and the neutrino mass hierarchy~\cite{PDG}. The recently commissioned JUNO experiment, designed primarily to determine the neutrino mass hierarchy, can also measure the mixing parameters with high precision and has already produced impressive first results. Its recently reported value~\cite{JUNO},
$sin^2\theta_{12}= 0.3092 \pm 0.0087$,
improves the precision by a factor of 1.6 relative to the combination of all previous measurements, and is expected to stimulate renewed interest in the study of neutrino mixing. In this work, we investigate the implications of this result for neutrino mixing patterns.

 For three neutrinos, the number of mixing parameters depends on whether neutrinos are Dirac or Majorana particles. For Dirac neutrinos, the lepton mixing matrix is described by three mixing angles, $\theta_{12}$, $\theta_{23}$, and $\theta_{13}$, and one CP-violating phase $\delta$ in the Particle Data Group (PDG) parameterization~\cite{PDG}. For Majorana neutrinos, there are two additional phases, called Majorana phases, which do not affect oscillation-related observables.
 
 Before the measurement of a nonzero $\theta_{13}$, one of the most popular mixing patterns was the tri-bimaximal mixing matrix $V_{TB}$~\cite{tri-bimaximal},
 \begin{eqnarray}
 	V_{TB} =
 	\left(
 	\begin{array}{ccc}
 		\sqrt{\frac{2}{3}} & \frac{1}{\sqrt{3}} & 0 \\
 		-\frac{1}{\sqrt{6}} & \frac{1}{\sqrt{3}} & \frac{1}{\sqrt{2}} \\
 		-\frac{1}{\sqrt{6}} & \frac{1}{\sqrt{3}} & -\frac{1}{\sqrt{2}}
 	\end{array}
 	\right).
 \end{eqnarray}
 This pattern was widely studied because of its simplicity, and many theoretical efforts were devoted to realizing it, including models based on the $A_4$ flavor symmetry~\cite{A4}.

 Since the discovery of a nonzero $\theta_{13}$ in 2012 by the Daya Bay neutrino experiment, the simple and elegant tri-bimaximal mixing pattern has to be modified. Three types of mixing patterns, which can be viewed as minimal modifications, have been proposed. They are obtained by rotating two of the neutrino generations in the basis where the charged leptons are already diagonalized so that the new mixing matrix becomes~\cite{hezee}
 $V_i = V_{TB} U_i$, with
 \begin{eqnarray}
 	U_a = \left ( \begin{array}{ccc}
 		\cos\tau& \sin\tau e^{i\alpha}&0\\
 		-\sin\tau e^{-i\alpha}& \cos\tau&0\\
 		0&0&1
 	\end{array}\right ), \;\;
 	U_b = \left ( \begin{array}{ccc}
 		\cos\tau&0& \sin\tau e^{i\alpha}\\
 		0&1&0\\
 		-\sin\tau e^{-i\alpha}& 0&\cos\tau
 	\end{array}\right ), \;\;
 	U_c = & \left ( \begin{array}{ccc}
 		1&0&0\\
 		0&\cos\tau& \sin\tau e^{i\alpha}\\
 		0&-\sin\tau e^{-i\alpha}& \cos\tau
 	\end{array}\right ).
 \end{eqnarray}
 Latter, we will use $c=\cos\tau$ and $s=\sin\tau$. These three modifications are minimal in the sense that they introduce the fewest additional parameters while also allowing for potentially CP-violating effects.
 
 For $U_a$, the phase $\alpha$ can be absorbed into the neutrino fields and therefore has no physical effect. The resulting mixing matrix $V_a$ keeps the third column of the tri-bimaximal matrix unchanged, implying
 $V_{e3}=0$.
 This mixing pattern is therefore not viable because the measured $\theta_{13}$ is nonzero.
 
 For $U_b$, we obtain
 \begin{eqnarray}
 	V_b = \left ( \begin{array}{ccc}
 		\frac{2c}{\sqrt{6}}& \frac{1}{\sqrt{3}}&\frac{2s}{\sqrt{6}}, e^{i\alpha}\\
 		-\frac{c}{\sqrt{6}} - \frac{s}{\sqrt{2}} e^{-i\alpha}& \frac{1}{\sqrt{3}}&\frac{c}{\sqrt{2}} - \frac{s}{\sqrt{6}} e^{i\alpha}\\
 		-\frac{c}{\sqrt{6}} - \frac{s}{\sqrt{2}} e^{-i\alpha}& \frac{1}{\sqrt{3}}&-\frac{c}{\sqrt{2}} - \frac{s}{\sqrt{6}} e^{i\alpha}
 	\end{array}\right ),
 \end{eqnarray}
 which predicts $V_{e2}=1/\sqrt{3}$.
 This was roughly in agreement with data within $2\sigma$ priori to the JUNO new result.
 It also predicts the CP-violating phase $\delta = \pm\pi/2$, with $\delta$ to be close to $-\pi/2$ favored by the inverted mass hierarchy. This mixing pattern has therefore attracted attention and has been studied previously, and it can be naturally derived from $A_4$ flavor-symmetry–based theories~\cite{heli}.
 
 Because $V_{e2} = \sin\theta_{12}\sqrt{1-\sin^2\theta_{13}}$, the newly measured JUNO value $\sin^2\theta_{12} = 0.309 \pm 0.0087$ implies that the predicted value $V_{e2}^2 = 1/3$ is now about $3.5\sigma$ away from the measurement. This mixing pattern is therefore at the edge of being ruled out. More precise data from future JUNO measurements may determine whether it can be firmly excluded. This makes it desirable to search for an alternative mixing pattern.

For $V_c$, we have
\begin{eqnarray}
	V_c = \left ( \begin{array}{ccc}
		\frac{2}{\sqrt{6}} & \frac{c}{\sqrt{3}} & \frac{s}{\sqrt{3}} e^{i\alpha} \\
		-\frac{1}{\sqrt{6}} & \frac{c}{\sqrt{3}} - \frac{s}{\sqrt{2}} e^{-i\alpha} & \frac{c}{\sqrt{2}} + \frac{s}{\sqrt{3}} e^{i\alpha} \\
		-\frac{1}{\sqrt{6}} & \frac{c}{\sqrt{3}} + \frac{s}{\sqrt{2}} e^{-i\alpha} & -\frac{c}{\sqrt{2}} + \frac{s}{\sqrt{3}} e^{i\alpha}
	\end{array}\right).
\end{eqnarray}
This modification keeps the first column of the tri-bimaximal mixing matrix.

In this case, $V_{e1} = 2/\sqrt{6}$, $V_{\mu 1} = V_{\tau1}=-1/\sqrt{6}$, and 
\begin{eqnarray}
	|V_{e2}|^2 = \frac{c^2}{3}\;, \;\;|V_{e3}|^2 =  \frac{s^2}{3}\;,\;\;
	|V_{\mu 3}|^2 = \frac{c^2}{2} + \frac{s^2}{3} + \frac{2 c s \cos\alpha}{\sqrt{6}}\;, \;\;
	|V_{\tau 3}|^2 = \frac{c^2}{2} + \frac{s^2}{3} - \frac{2 c s \cos\alpha}{\sqrt{6}}\;.
\end{eqnarray}
These equations can be used to determine the values of $c$ and $\cos\alpha$ from the measured $\theta_{12}$ and $V_{\mu 3}$.
Also $\theta_{12}$ and $\theta_{13}$ are not independent, but related by, 
\begin{eqnarray}
	\sin^2\theta_{12} = {1-3\sin^2\theta_{13}\over 3 (1-\sin^2\theta_{13})}\;.
\end{eqnarray}

We now study the implications of this mixing pattern. 
The appearance of $c$ in $V_{e2}$ provides a way to bring $V_{e2}$ closer to the JUNO-measured value.
The CP-violating phase $\delta$ can be related to $c$ and $\alpha$ by comparing the rephasing-invariant quantity~\cite{JJ}, $Im(V_{e1}V_{e2}^* V^*_{\mu1} V_{\mu 3})$, in the PDG parameterization and in $V_c$. We have
\begin{eqnarray}
	\sin\delta_{CP} = - \frac{(1 - s^2/3)\sin\alpha}{2 \sqrt{(c^2/2 + s^2/3)^2 - 2 c^2 s^2 \cos^2\alpha / 3}}.
\end{eqnarray}

One can also relate $\cos\delta$ with $c$ and $\alpha$ by comparing $V_{\mu 2}$ and $V_{\tau 2}$ in the PDG and $V_c$ parameterizations:
\begin{eqnarray}
	\cos\delta_{CP} = \frac{\sqrt{6}}{4 s c |V_{\mu 3}|| V_{\tau 3}|} \left[ \frac{1}{2} \left(1 - \frac{s^2}{3} \right)^2 - c^2 |V_{\tau 3}|^2 - \frac{2}{3} s^2 |V_{\mu 3}|^2 \right].
\end{eqnarray}

 To fix the mixing matrix completely, one needs another input values to detetmined the phase $\alpha$. 
 We use the measured $|V_{\mu3}|^2 = 1/2$ as the central vlue with an error of 0.02 for illustration. This value is compatible with current data~\cite{PDG}. 
 
 Eq.(6) one would predict, $\sin^2\theta_{12} = 0.3183\pm 0.0042$~\cite{PDG}. This is in agreement withthe JUNO measurement $\sin^2\theta_{12} = 0.3092\pm 0.0087$ within 1$\sigma$.  In our later discussion we will use  the more precisely measured $\theta_{13}$ to detemine the value of $c$.  We obtain
 \begin{eqnarray}
 	c=0.952\pm 0.013\;,\;\;\cos\alpha = 0.066\pm 0.087\;.
 \end{eqnarray}

We also obtain $\sin\delta_{CP} \approx -\sin\alpha$ whose size is 0.998. We also find that  $\cos\alpha$ is determed to be positive indicating that $\delta$ is in the first or fourth quadrant. However, the sign of $\sin\delta_{CP}$ is still not determined.
 The sign of $\sin\delta_{CP}$ depends on the sign of $\alpha$. 
 If taking $\alpha$ to be positive, one would obtain $\sin\delta_{CP} = -0.998 \pm 0.3$ which is favored by the inverted neutrino mass hierarchy. Other otherwise, $\sin\delta_{CP} = 0.998 \pm 0.3$.  This is not favored by neither inverted herarchy and normal neutrino mass hierarchies.
 
 In conclusion, we have studied the minimally modified tri-bimaxl mixing and find a mixing parttern that keeps the first column of the tri-bimaximal mixing matrix fits data well within 1$\sigma$ level and predicts CP violating phase $\delta_{CP}$ about $-\pi/2$. This can be tested by future experimental data, such as those will come from DUNE experiment.

 \section*{Acknowledgments}

This work was supported by the Fundamental Research
Funds for the Central Universities, by the National Natural Science Foundation of the People’s
Republic of China (Nos. 12090064, 11735010, 12205063, 11985149, 12375088, and W2441004),
and by MOST 109-2112-M-002-017-MY3.

\end{document}